\documentclass[aps,prx,preprint,groupedaddress,amssymb,amsmath,floatfix]{revtex4-1} 

\usepackage{graphicx}
\usepackage{bm}
\usepackage{cleveref}
\usepackage{epstopdf}

\usepackage{subfig}
\usepackage{color}

\begin{document}
\title{Highly Valley-Polarized Singlet and Triplet Interlayer Excitons in van der Waals Heterostructure}
\author{Long Zhang$^1$} \email{lonzhang@umich.edu}
\author{Rahul Gogna$^2$, G. William Burg$^3$, Jason Horng$^1$, Eunice Paik$^1$, Yu-Hsun Chou$^1$, Kyounghwan Kim$^3$, Emanuel Tutuc$^3$}
\author{Hui~Deng$^{1,2}$} \email{dengh@umich.edu}
\small{ }
\address{$^1$ Physics Department, University of Michigan, 450 Church Street, Ann Arbor, MI 48109-2122, USA}
\address{$^2$ Applied Physics Program, University of Michigan, 450 Church Street, Ann Arbor, MI 48109-1040, USA}
\address{$^3$ Microelectronics Research Center, Department of Electrical and Computer Engineering, The University of Texas at Austin, Austin, Texas 78758, United States
}

\begin{abstract}
     Two-dimensional semiconductors feature valleytronics phenomena due to locking of the spin and momentum valley of the electrons. 
     However, the valley polarization is intrinsically limited in monolayer crystals by the fast intervalley electron-hole exchange.
     Hetero-bilayer crystals have been shown to have a longer exciton lifetime and valley depolarization time. 
     But the reported valley polarization was low; the valley selection rules and mechanisms of valley depolarization remains controversial. 
     Here, we report singlet and brightened triplet interlayer excitons both with over 80\% valley polarizations, cross- and co-polarized with the pump laser, respectively. This is achieved in WSe$_{2}$/MoSe$_{2}$ hetero-bilayers with precise momentum valley alignment and narrow emission linewidth. The high valley polarizations allow us to identify the band minima in a hetero-structure and confirm unambiguously the direct band-gap exciton transition, ultrafast charge separation, strongly suppressed valley depolarization. Our results pave the way for
     using semiconductor heterobilayers to control valley selection rules for valleytronic applications.
\end{abstract}


\maketitle

\section{Introduction}
Monolayer transition metal dichalcogenide crystals (TMDCs) feature strong intrinsic spin orbital coupling (SOC) and broken inversion symmetry. Consequently, excitons from opposite momentum valleys couple with circularly polarized lights with opposite helicities \cite{mak_control_2012,xiao_coupled_2012}, enabling novel valleytronic phenomena and applications \cite{zhang_electrically_2014,wang_colloquium_2018}.
However, strong inter-valley scattering due to the electron hole (e-h) exchange interaction leads to rapid valley depolarization on a picosecond scale \cite{yu_dirac_2014,glazov_exciton_2014,wang_valley_2014,lagarde_carrier_2014,zhu_exciton_2014}, posing intrinsic limits on valley polarization (VP) of monolayer excitons.
Alternatively, interlayer excitons in hetero-bilayers may enable high VP since the e-h exchange interaction becomes suppressed due to reduced electron-hole wavefunction (\cref{fig:Schematic}a-b) \cite{rivera_interlayer_2018,jin_ultrafast_2018,mak_opportunities_2018}. Recent theoretical work also suggests the possibility to control the exciton states and valley selection rules through moir\'e superlattices in hetero-bilayers \cite{yu_moire_2017,wu_theory_2018}.  However, the reported VP of interlayer excitons has been low, comparable to that of intralayer ones and varying between 10$\%$ to 35$\%$, \cite{rivera_valley-polarized_2016,miller_long-lived_2017,hanbicki_double_2018,ciarrocchi_polarization_2019,tran_moire_2018}. Inconsistent valley selection rules and different mechanisms for the low VP have been proposed, such as slow charge separation \cite{rivera_valley-polarized_2016}, formation of indirect bandgap transitions and compromised optical selection rules even in rotationally aligned bilayers \cite{hanbicki_double_2018,ciarrocchi_polarization_2019}, and mixing among different mini-bands in a moir\'e lattice \cite{tran_moire_2018}. In many of these works, the interlayer exciton emission shows an inhomogeneous broadening of 20-50 meV, which may have masked the valley selection rules of individual exciton states and yielded inconsistent and controversial results. (see the Supplementary materials)

In this work, using rotationally aligned WSe$_{2}$/MoSe$_{2}$ hetero-bilayers with hexagon-Boron Nitride (hBN) encapsulation, we observe narrow linewidth of 6~meV, thereby resolving spin singlet and brightened spin triplet excitons with very high VPs over 80\%, in opposite helicities. 
Compared to previous work \cite{rivera_valley-polarized_2016,miller_long-lived_2017,hanbicki_double_2018,ciarrocchi_polarization_2019,tran_moire_2018,seyler_signatures_2018}, the high VP enables us to identify ambiguously the $H_h^h$ atomic registry as the exciton band minimum with a direct-bandgap transition, where the triplet exciton, which is dark in monolayers, became bright and forms the interlayer exciton ground state. The high VP also confirms the preservation of valley selection rules, ultrafast charge transfer and suppressed intervalley scattering in hetero-bilayers.  
These results pave the way for using heterostructures to control spin and charge dynamics and manipulate the optical selection rules .

\section{Results}
An optical image of the hBN-encapsulated WSe$_{2}$/MoSe$_{2}$ hetero-bilayer is shown in \cref{fig:Sample}a. The twist angle between WSe$_{2}$ and MoSe$_{2}$ is measured to be 58.7$^\circ$ $\pm$0.7$^\circ$ by angle-dependence of the second harmonic generation from the two monolayers and from the bilayer \cite{jiang_valley_2014,hsu_second_2014} (see Supplementary Materials for details).
The intra-layer and inter-layer exciton resonances are clearly identified in reflection contrast and photoluminescence (PL) measurements (\cref{fig:Sample}b). The absence of interlayer excitons in the reflectance contrast spectrum is expected because of smaller oscillator strengths due to the reduced e-h spatial overlap.
In PL, however, the intralayer exciton emission is quenched while the interlayer excitons is much brighter. This observation suggests fast separation of the electron and hole into the two stacked monolayers compared to the intralayer exciton recombination. It also confirms the two monolayer are aligned close to (multiples of) $60^\circ$, so that the momentum mismatch between electron and hole is small \cite{nayak_probing_2017}. We can still observe weak intralayer exciton emission from MoSe$_{2}$, suggesting slower hole transfer between the two layers.

Importantly, the linewidth of the interlayer exciton emission is only about 6~meV, showing greatly reduced inhomogeneous broadening thanks to encapsulation, which enables us to resolve individual exciton transitions. In contrast, in hetero-bilayers without hBN encapsulation and with a larger linewidth of 40 meV, VP of the interlayer exciton emission remained less than $35\%$; the helicity varied from sample to sample rather than reflecting the any underlying individual transitions (see Supplemental Figure 2).
%
%

We further examine the properties of the interlayer excitons and their dependence on the intralayer excitons via PL excitation spectroscopy. We scan a continuous-wave excitation laser with $\sigma^{+}$ circular polarization across the WSe$_{2}$ and MoSe$_{2}$ intralayer exciton resonances, while monitoring the interlayer exciton emission with co-circular ($\sigma^{+}/\sigma^{+}$) and counter-circular ($\sigma^{+}/\sigma^{-}$) polarizations, as shown in the left and right columns, respectively, in \cref{fig:PLE}a.

\begin{figure*}[t]
	\includegraphics[width=\linewidth]{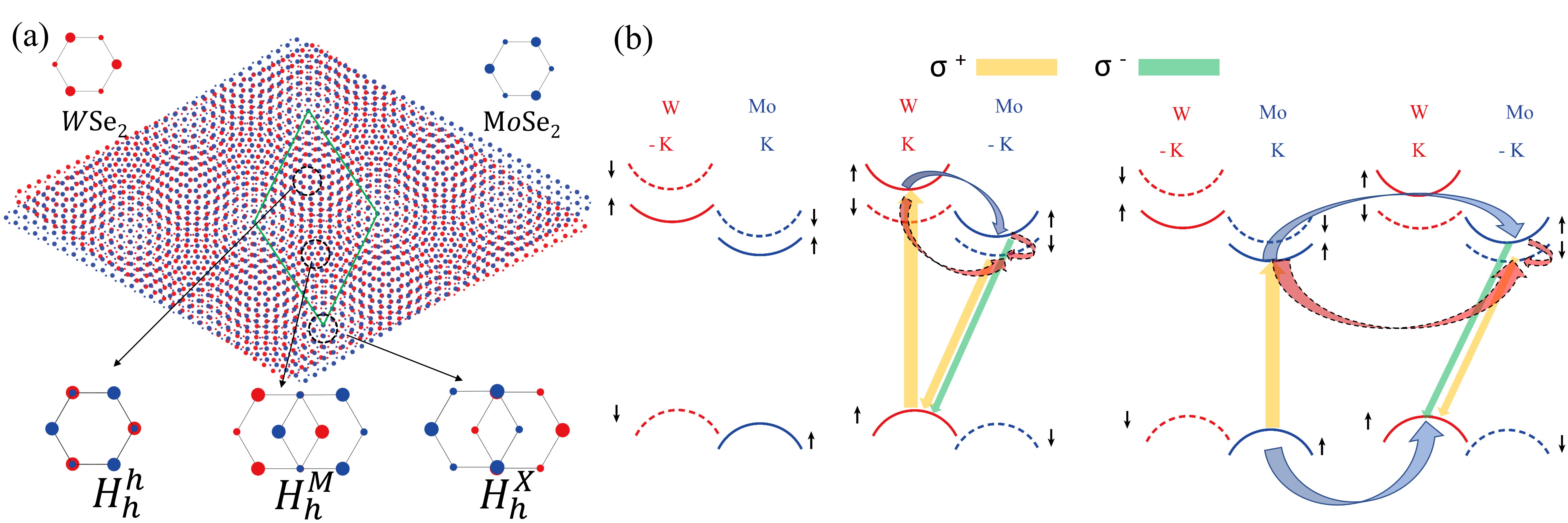}
	\caption{Moir\'e lattice and interlayer excitons dynamics in the hetero-bilayers. (a) Illustration of a moir\'e lattice formed in a WSe$_{2}$/MoSe$_{2}$ hetero-bilayer, when the twist angle between the monolayer lattices is close to $60^{\circ}$. Within the moir\'e unite cell, the three high symmetry points are labeled by the black circles, and the lattice stacking orders are shown at the bottom. The $H^{\mu}_{h}$ donates an H type stacking order with the $\mu$ site of the MoSe$_{2}$ lattice vertically aligned with the hexagon center (h) of the WSe$_{2}$ lattice \cite{yu_moire_2017}. (b),(c) Schematic illustration of the carrier transfer, relaxation and radiative recombination processes when the bilayer is pumped at the intralayer excitons at the K valley of WSe$_{2}$ or MoSe$_{2}$, respectively. The Black arrows indicate the spin configurations in the conduction and valence bands. Solid (dashed) lines indicate the spin up (down) state. The solid (dashed curved) arrows indicate the spin-conserving (spin-flipping) carrier transfer processes. The orange (green) straight arrows indicate transitions coupled to light with $\sigma^+$ ($\sigma^-$) polarized light, respectively.
}
    \label{fig:Schematic}
\end{figure*}

\begin{figure*}[t]
	\includegraphics[width=\linewidth]{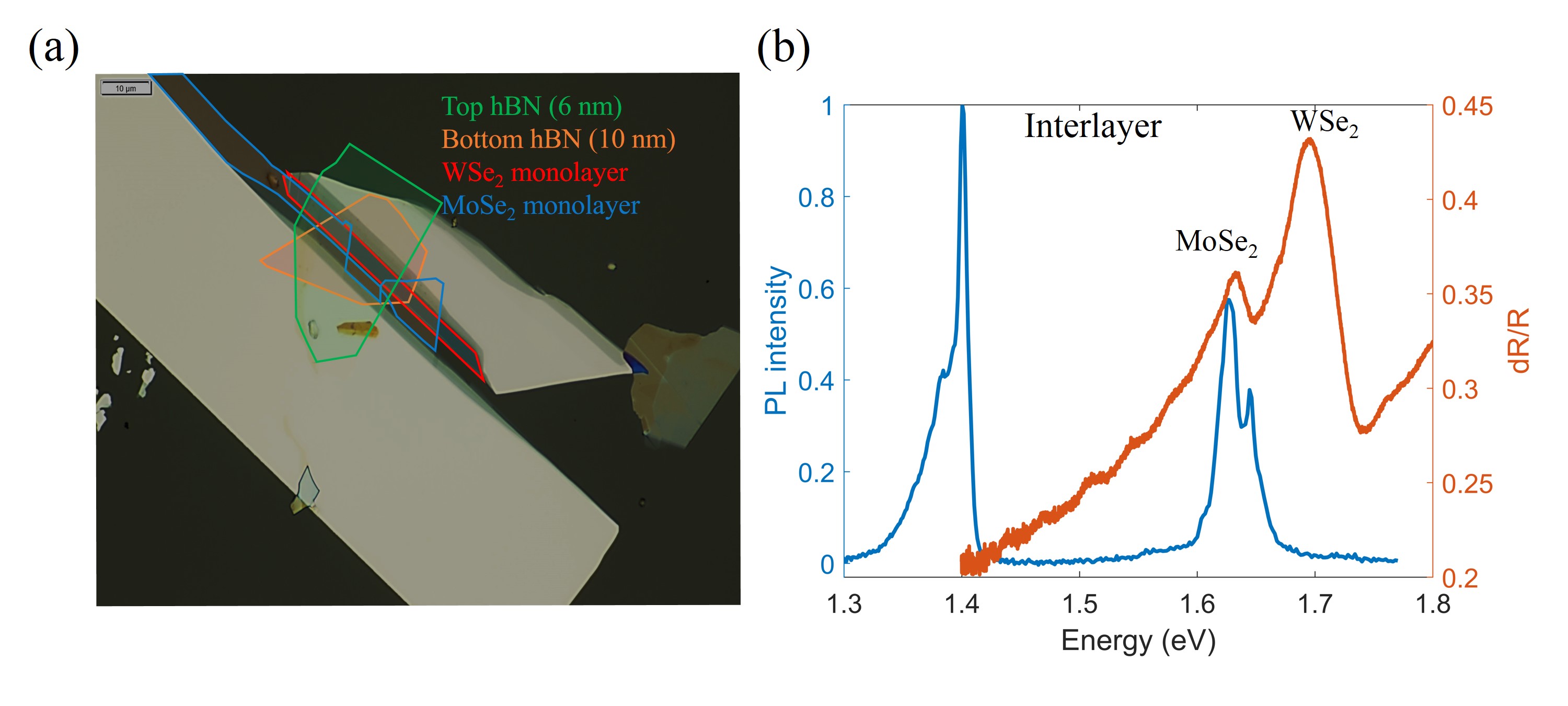}
\caption{ Interlayer excitons in WSe$_{2}$/MoSe$_{2}$ heterobilayer
(a)Optical mircograph of the hBN encapsulated WSe$_{2}$/MoSe$_{2}$ heterostructure. The solid lines indicate the contours of the different layers. (b) Reflection contrast (red) and PL (blue) spectra of the heterostructure. The PL is obtained with a continuous-wave pump laser at 1.95 eV.
}
\label{fig:Sample}
\end{figure*}

\begin{figure*}[t]
  \includegraphics[width=\linewidth]{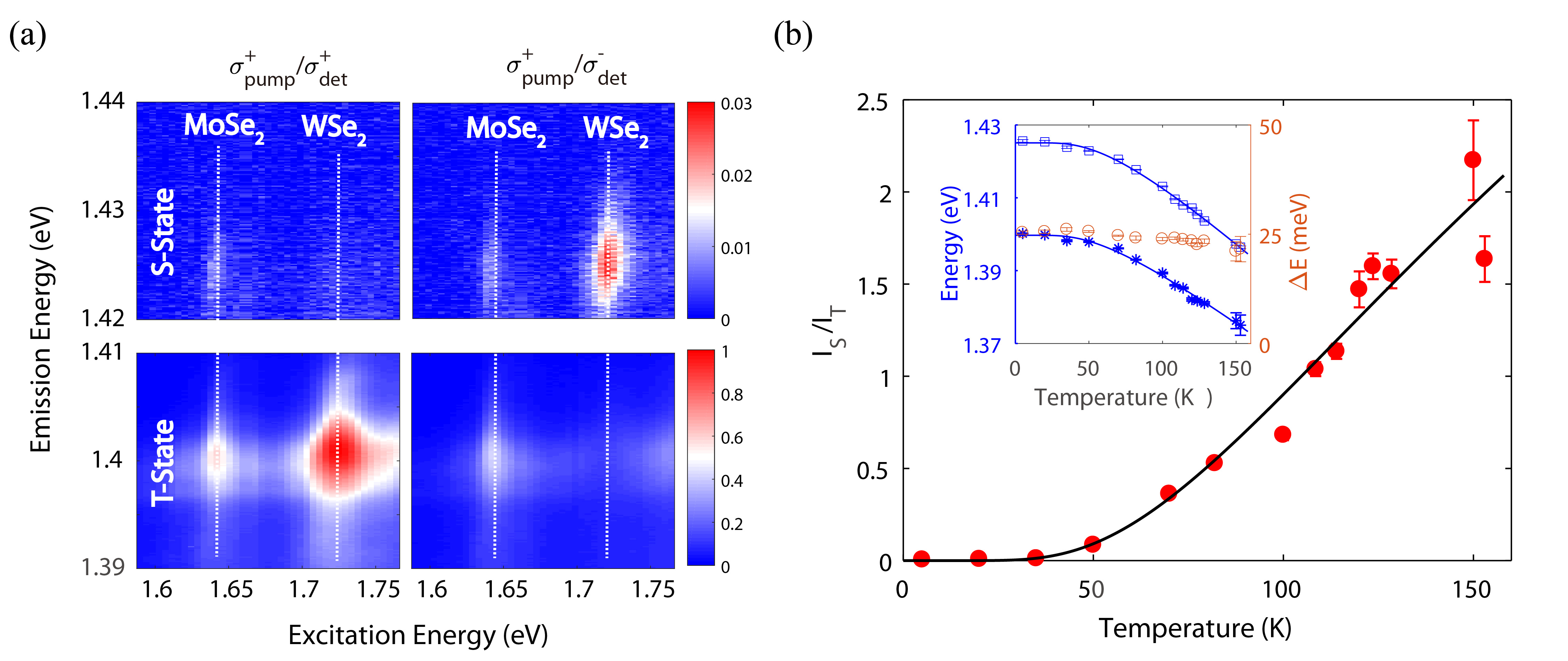}
\caption{Valley-polarized singlet and triplet interlayer exciton emission by photoluminescence excitation spectroscopy.
(a) Polarization-resolved emission spectra as a function of the excitation laser energy. The color represents the emission intensity. The left (right) column shows the spectra co-polarized (cross-polarized) with the $\sigma^+$ pump. The top (bottom) row shows emission from the S-state (T-state), or singlet (triplet) interlayer excitons, evidently cross-polarized (co-polarized) with the pump. The white dashed lines indicate the MoSe$_{2}$ and WSe$_{2}$ A exciton resonances. (b) Temperature dependence of the PL intensity ratio of the S-state vs. the T-states when pumped at the WSe$_{2}$ A exciton resonances (filled red circles). The black solid line is the fit based on the Boltzman distribution.  The inset shows the temperature dependence of the energies of the S-state and T-state (blue squares and stars). The blue solid lines are the fits by the typical temperature dependence of the semiconductor band gap. The energy difference between the two states are shown by the open orange circles and stays approximately constant with temperature.
}
	\label{fig:PLE}
\end{figure*}

\begin{figure*}[t]
  \includegraphics[width=\linewidth]{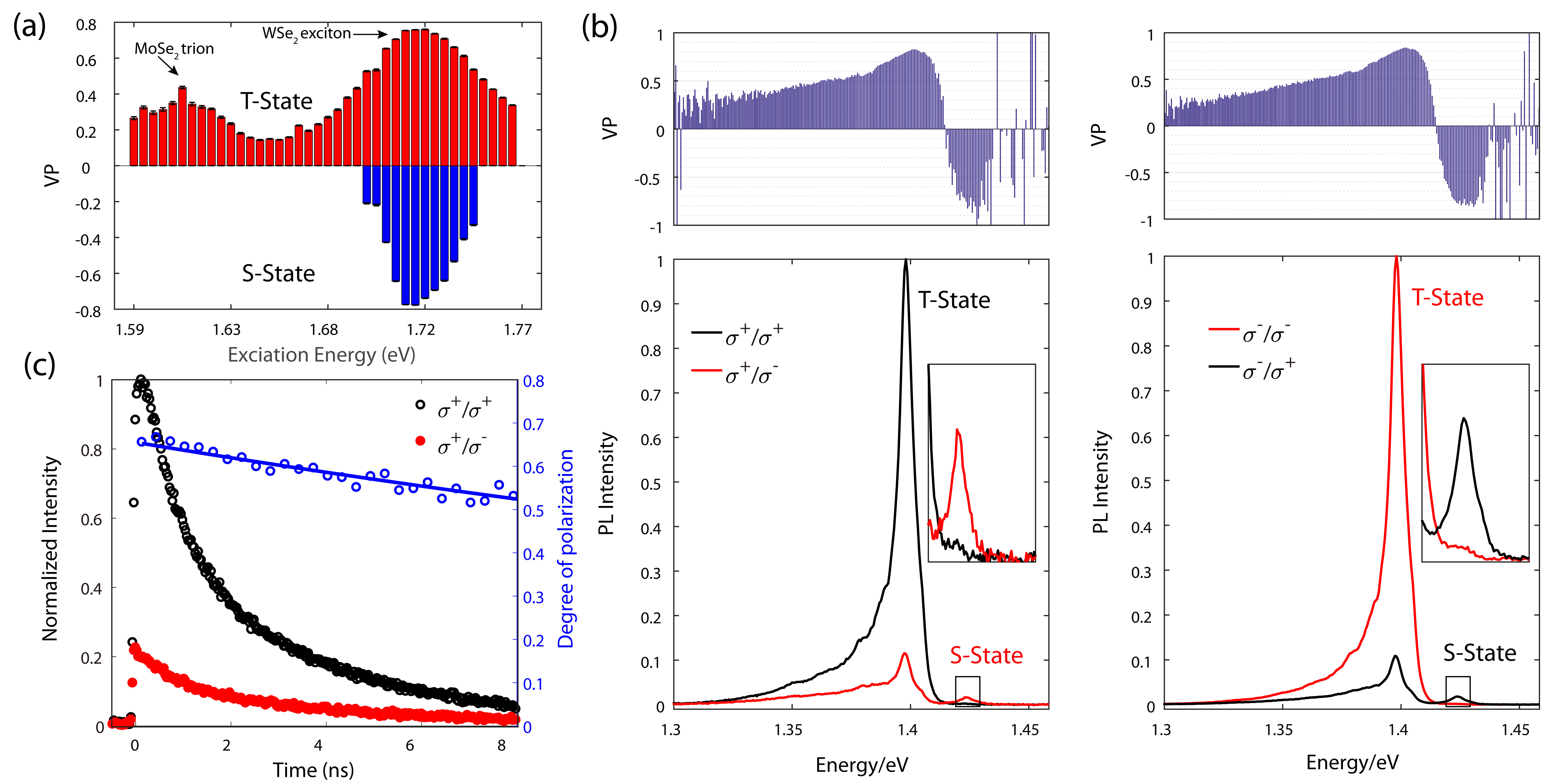}
	\caption{
High valley polarizations of the interlayer excitons.
(a) VPs of the S-state and T-state emission as a function of the excitation energy. The higher VP when pumped at MoSe$_{2}$ trion resonance compared to neutral exciton is consistent with previous report of much longer trion valley lifetime than neural exciton \cite {singh_long-lived_2016}. (b) Bottom panels show the polarization-resolved PL spectra of the heterostructures when pumped at the WSe$_{2}$ A exciton resonances, showing a large difference between the $\sigma^{+}$ (black) and $\sigma^{-}$ (red) polarized components. The corresponding VP is shown in the top panel. Two peaks, the S-state and T-state, are clearly identified with opposite helicities. The insets are zoom-in of the spectrum region of the S-state, as marked by the rectangles. (c) Time evolution of the co-polarized (black circles) and cross-polarized (red circles) PL of the interlayer excitons and the corresponding VP (blue circles). The solid line is a single exponential fit of the VP decay with a fitted decay time of $\tau=32.5\pm4.1$~ns.
}
    \label{fig:TRPL}
\end{figure*}

The data clearly show two interlayer exciton states with opposite and high VPs: a strong emission peak at 1.400~eV co-polarized with the pump, label as the T-state (bottom row), and a much weaker emission peak at 1.425 eV, cross-polarized with the pumped, label as the S-state (top row).
Both states show significantly enhanced emission intensities and VPs as the excitation wavelength coincide with the WSe$_{2}$ and MoSe$_{2}$ intralayer exciton resonances (\cref{fig:PLE}a and \cref{fig:TRPL}a), which confirms that T and S states are interlayer excitons with VPs inherited from the intralayer excitations. The two states are separated by 25~meV, corresponding to the MoSe$_{2}$ conduction band splitting associated with SOC \cite{larentis_large_2018,liu_electronic_2015}, which suggests that the higher-energy S-state is the bright singlet interlayer exciton, while the lower-energy T-state is the brightened triplet interlayer exciton (\cref{fig:Schematic}b,c).

To understand the intensity difference between the S-state and T-state, and further confirm their origin, we measure the temperature dependence of the emission from 5~K to 150~K with a $\sigma^+$ excitation laser at 1.72 eV. The spectrum is shown in supplementary material. The resonance energies of both states redshift as temperature increases, which are well described by standard temperature dependence of semiconductor bandgap: $E_g(T)=E_g(0)-S\hbar\omega[coth(\hbar\omega/2k_{B}T-1)]$, where $E_g(0)$ is the exciton resonance energy at T=0 K, S is the dimensionless coupling constant, and $\hbar\omega$ is the average phonon energy\cite{odonnell_temperature_1991}. From the fits, we extract for triplet and (singlet) state, the $E_g(0)=1.400\pm0.001  (1.425\pm 0.001)eV$, $S=1.9\pm0.3(2.2\pm0.2)$, $\hbar\omega=15.3\pm3.1 meV$ for both. The separation $\Delta E$ between the two states stays consistently between 22~meV and 25~meV (inset of \cref{fig:PLE}b).
With a constant $\Delta E$,  the population in the two states should follow the Boltzman distribution of the population at equilibrium; the ratio of their total emission intensities $I_{S}/I_{T}$ is then given by:
\begin{equation}\label{eq:Int-T}
  I_{S}/I_{T}= (\tau_T/\tau_{S}) e^{(-\Delta E/k_{B}T)},
\end{equation}
where $\tau_{S,R}$ is the decay time of the S- and T-state.  The equation \cref{eq:Int-T} fits the data very well. From the fit, we obtain $\Delta E= 24.1$~meV$\pm 3.6$~ meV, consistent with the measured S- and T-state separation as well as the conduction band splitting of MoSe$_{2}$ \cite{larentis_large_2018,liu_electronic_2015}. The fitted ratio $\tau_T/\tau_{S}=14.1 \pm 4.7$ suggests the singlet state recombines much faster than the triplet state, also consistent with the calculation \cite{yu_brightened_2018}.

To understand the VPs of the singlet and triplet exciton emission, we analyze the optical selection rules for the heterostructure,  as illustrated in \cref{fig:Schematic}. As the two monolayers with a lattice mismatch of $\delta a$ are stacked together with a $60^\circ$ twist angle, a moir\'e superlattice is formed with a period of $a/\delta_a$ (\cref{fig:Schematic}a). The atoms in the two monolayers are displaced from each other except at the three special positions in the moir\'e super-cell, as illustrated in \cref{fig:Schematic}a. At these points, the atomic registry recovers the three fold rotation symmetry, and therefore, the VPs are restored for the excitonic transitions.
The three points also correspond to potential extrema in the superlattice potential, as was predicted by density function theory (DFT) calculation and confirmed by the scanning tunneling spectroscopy (STM) in MoS$_{2}$/WSe$_{2}$ hetero-bilayers \cite{wu_theory_2018,yu_moire_2017, zhang_interlayer_2017}.

At the $H_{h}^{h}$ registry (\cref{fig:Schematic}a), the exciton singlet (triplet) state couples to circularly polarized light with the opposite (same) helicity as that of the intralayer exciton in the same valley \cite{yu_brightened_2018}, which fully agrees with our observation. At the $H_{h}^{X}$ and $H_{h}^{M}$ registries, the singlet and triplet excitons couple to light with either the opposite helicity than as observed or with an out-of-plane polarization. The measured high VP suggests the emission comes from excitons localized at the $H_{h}^{h}$ registry. The relatively high T-state emission intensity compared to the S-state or intralayer excitons suggests that the $H_{h}^{h}$ registry corresponds to the potential minimum in the moir\'e lattice.

The high VPs and their dependence on the excitation wavelength also shed light on the charge and spin relaxation processes in the heterostructure.
We first analyze more closely the VP of the interlayer excitons.
The VP depends sensitively on the excitation energy. 
As shown in \cref{fig:TRPL}a, 
the S- and T-state, exhibiting positive and negative helicity respectively, both reaching maximum absolute values of VP of $0.8$ when the excitation laser is resonant with the WSe$_2$ intralayer exciton energy. 

With the pump fixed at the WSe$_2$ intralayer exciton resonance, we measure the interlayer exciton VP under both $\sigma^{+}$ (red lines) and $\sigma^{-}$ (black lines) polarized pumping (\cref{fig:TRPL}b). With both pump polarizations, we measure bright co-polarized T-state emission and weaker cross-polarized S-state emission, both with the absolute values of VP up to $0.8$. These pronounced features including two interlayer exciton states with alternate helicities and high VP, are consistently observed across the whole sample, confirming the high uniformity of the heterostructure (see Supplementary Material). Similar results are also reproduced in a few other high quality MoSe$_{2}$/WSe$_{2}$ hetero-bilayers (see Supplementary Material). 

The very high VP measured is possible only if we have a high intralayer VP that is well preserved by the interlayer excitons. Preserving the intralayer VP requires rapid relaxation from intralayer excitons to the interlayer exciton states at the high symmetry points, and slow valley depolarization of the intervalley excitons compared to their recombination time.
As illustrated in \cref{fig:Schematic}b, resonantly excited WSe$_2$ intralayer excitons have a high initial VP. Before they can recombine or scatter to the opposite valley on the pico-second time scale, the conduction band electrons rapidly transfer to lower energy states in the overlapping MoSe$_{2}$ layer, conserving spin and momentum \cite{hong_ultrafast_2014, schaibley_directional_2016,jin_ultrafast_2018}, followed by energy relaxation to the band edge. The valence band hole is already at the band edge and stays in WSe$_2$. Once the electron and hole are separated into two layers, the exchange interaction and therefore the valley-depolarization is suppressed. The electrons can thermalize to the lower conduction band with flipped spins. This process can happen efficiently in heterobilayers because of the conduction-band spin hybridization in the absence of the mirror symmetry \cite{yu_brightened_2018}. The electrons and holes, both in the same valley, form spin singlet and triplet states at moir\'e potential minima at the $H_{h}^{h}$ registries and radiatively recombine, emitting light with VP inherited from intravalley excitons. 

The interlayer exciton VP is lower when pumped at the MoSe$_{2}$ exciton resonance. It is expected due to a lower initial intralayer exciton VP in MoSe$_{2}$ \cite{wang_polarization_2015,macneill_breaking_2015}. The formation of the  interlayer excitons may also be slower as it requires both inter-valley scattering and flipping of the electron spin (see illustration in \cref{fig:Schematic}c). 

To compare the intervalley exciton depolarization time with the recombination time, we perform polarization resolved and time-resolved PL using a femto-second pulsed excitation laser resonant with WSe$_2$ intralayer exciton.  As shown in \cref{fig:TRPL}c,  fitting the data with single exponential function, we observe a short PL decay time $\tau_r=2.3\pm0.3$~ns, but a much longer valley depolarization time $\tau_v=32.5\pm4.1$~ns. The VP $\rho$ of the integrated PL can be related to the initial VP $\rho_{0}$ by \cite{lagarde_carrier_2014,wang_valley_2014}:
\begin{equation}\label{eq:rho_0}
  \rho=\frac{\rho_{0}}{1+2\frac{\tau_r}{\tau_v}}=(87.6\%\pm 7.9\%)\rho_0.
\end{equation}
Therefore $87.6\%$ of the initial VP is retained for interlayer excitons. The measured VP of $81.3\%\pm 4.9\% $ suggests an initial $\rho_0^{inter}=92.8\% \pm 8.9\%$ for the interlayer excitons. Assuming a near-perfect initial VP of $\rho_0^{intra}$ for the resonantly pumped WSe$_2$ excitons, substituting $\rho_0^{inter}$ and $\rho_0^{intra}$ into \cref{eq:rho_0}, we deduce that the initial charge separation takes place more than ten times faster than the intralayer exciton valley depolarization rate.


\section{Conclusions}
In summary, we observe emission from two interlayer exciton states with very high VPs and opposite helicities in WSe$_2$/MoSe$_2$ hetero-bilayers with a $60^{\circ}$ twist angle. The high VP and short lifetimes confirm they are direct bandgap excitons as opposed to indirect ones between the $Q$ and $\Gamma$ valleys \cite{ciarrocchi_polarization_2019,hanbicki_double_2018}. We identify the two exciton states as spin singlet and triplet excitons localized at the $H^h_h$ atomic registry based on their helicities \cite{yu_brightened_2018}, energy separation and temperature dependence of the emission intensities. The relative oscillator strengths of the two states are also obtained from the temperature dependence, which is consistent with the singlet and triplet assignment \cite{yu_brightened_2018} rather than two minibands in a moir\'e lattice \cite{wu_theory_2018,yu_moire_2017}.
We are able to identify and measure highly valley-polarized singlet and tripilet excitons thanks to the very small inhomogeneous broadening of about 6~meV with hBN encapsulation. This is in sharp contrast to other reports in the literature with an inhomogeneous linewidth of 20-50~meV \cite{rivera_valley-polarized_2016,miller_long-lived_2017,tran_moire_2018} and VPs below 35\% when different exciton states cannot be clearly resolved.

The high VPs show rapid electron transfer between the two monolayers at timescales an order of magnitude shorter than the WSe$_2$ intralayer exciton depolarization time, and a long interlayer exciton valley depolarization time compared to the recombination time. A valley depolarization time of about 33~ns is measured, suggesting strongly suppressed inter-valley exchange interactions thanks to both electron-hole separation into the two monolayers and discretized spin singlet and triplet states.
%
The VP in these heterobilayers can be altered by using different twist angles and materials and can be further controlled by electric fields or strain \cite{yu_moire_2017}. A heterostructure moir\'e lattice may enable further control of spin orbital coupling and open doors to other novel topological states.


\section{Materials and Methods}
\noindent\textbf{Heterostructure fabrication.}  Monolayers of MoSe$_{2}$, WSe$_{2}$ and thin flakes of hBN are first mechanically exfoliated onto 300 nm SiO$_{2}$ on Si wafers. We then use the dry transfer method to pick up and stack up the crystals to create the heterostructure, with the MoSe$_{2}$ and WSe$_{2}$ armchair aligned under a microscope, followed by annealing in vacuum \cite{kim_van_2016}.

\noindent\textbf{Experimental setup}
For low temperature measurements, the sample is kept in a 4K cryostat (Montana Instrument).  The excitation and collection are carried out with a home-built confocal microscope with an objective lens with numerical aperture (NA) of 0.45. For reflection contrast measurement, white light from a tungsten halogen lamp is focused on the sample with beam size of 10 $\mu$m in diameter. The spatial resolution is improved to be 2 $\mu$m by using pinhole combined with confocal lens. For PL measurements,
a continuous wave Ti:sapphire laser (MSquared-Solstis, bandwidth $<$50~kHz, power held at 80~$\mu$W) is focused by the same objective with beam size of 2 $\mu$m. The signal is detected using a Princeton Instruments spectrometer with a cooled charge-coupled camera for time-integrated measurements. For  time-resolved measurements, we use a single photon detector synchronized with the laser with a time resolution below 200~ps. 

\section{Author Contributions}
  H.D., L.Z.conceived the experiment. G.W.B, L.Z. fabricated the device. L.Z., R.G. performed the measurements. L.Z. and H.D. performed data analysis. J.H., E.P., Y.C, K.K assisted the fabrication. H.D. and E.T. supervised the projects. L.Z and H.D. wrote the paper. All authors discussed the results, data analysis and the paper.
\section{Acknowledgment}
We thank Allan H. MacDonald, Fengcheng Wu, Wei Xie, Wencan Jin and Kai Chang for helpful discussions. All authors acknowledge the support by the Army Research Office under Awards W911NF-17-1-0312.
\bibliographystyle{nature3}
\bibliography{Reference}

\pagebreak

\end{document}